\documentstyle[amsfonts,11pt]{article}

\def\bbt{\bibitem}
\def\be{\begin{equation}}
\def\en{\end{equation}}
\def\ber{\begin{eqnarray}}
\def\enr{\end{eqnarray}}
\def\nmb{ \nonumber\\}
\def\d{\partial}

\def\ov{\over }
\def\tld{\tilde}

\def\sgm{\sigma}

\def\im{\imath}

\def\om{\omega}
\def\et{\eta}

\def\dlt{\delta}

\begin{document}
\rightline{Landau Tmp/06/02.}
\rightline{February 2002}
\vskip 2 true cm
\centerline{\bf FREE FIELD CONSTRUCTION OF D-BRANES}
\centerline{\bf IN $N=2$ SUPERCONFORMAL MINIMAL MODELS.}
\vskip 1.5 true cm
\centerline{\bf S. E. Parkhomenko}
\centerline{Landau Institute for Theoretical Physics}
\centerline{142432 Chernogolovka, Russia}
\vskip 0.5 true cm
\centerline{spark@itp.ac.ru}
\vskip 1 true cm
\centerline{\bf Abstract}
\vskip 0.5 true cm

\centerline{The construction of D-branes in $N=2$ superconformal minimal
models}
\centerline{based on free-field realization of $N=2$ super-Virasoro algebra
unitary modules is represented.}
\vskip 10pt

"{\it PACS: 11.25Hf; 11.25 Pm.}"


\smallskip
\vskip 10pt
\centerline{\bf0. Introduction}
\vskip 10pt

 The role of $D$-branes ~\cite{Pol1} in the description of certain
nonperturbative degrees of freedom of strings is by now well
established and the study of their dynamics has lead to many
new insights into string  and $M$-theory ~\cite{Wit},~\cite{Pol2}.
Much of this study was done in the large volume regime where
geometric techniques provide reliable information. The extrapolation
into the stringy regime usually requires boundary conformal field
theory (CFT) methods. In this approach $D$-brane configurations are
given by conformally invariant boundary states or boundary conditions.
However a complete microscopic description these configurations are well
understood only for the case of flat and toric backgrounds where
the CFT on the world sheet is a theory of free fields. Due to this
reason boundary state formalism for $D$-branes has been
subsequently developed and many calculations concerning the
scattering and couplings of
closed strings in a $D$-brane background have been given exactly
in ~\cite{VFPSLR}-~\cite{VFLL}, ~\cite{HINS}.

 The class of models of rational CFT gives the examples of curved
string backgrounds where the construction of the boundary states
leaving a whole chiral symmetry algebra unbroken can
be given in principle and the interaction of these states
with closed strings can be calculated exactly.
But in practice the calculation of closed string amplitudes in
general CFT backgrounds is
available only if the corresponding free field realization of
the model is known. Therefore, it is important to extend free
field approach to the case of rational models of CFT with a
boundary.

 This problem has been treated recently to the case of $SU(2)$
WZNW model in ~\cite{SP}, where the Wakimoto free field
realization of $sl(2)$ Kac-Moody algebra ~\cite{BFeld}-~\cite{GMOM} has
been used to boundary states construction.

 $N=2$ superconformal minimal models represent a subclass of
rational CFT which is most important in the string theory applications.
They are building blocks in Gepner models ~\cite{Gep} which give
an exact solution by CFT methods of the problem of string propagation on
Calabi-Yau manifolds. The $N=2$ minimal models
as suggested are fixed points of
$N=2$ supersymmetric Landau-Ginzburg (LG) models ~\cite{VW}, ~\cite{M}.
This rigorously justified suggestion gives a possibility to
describe string propagation on Calabi-Yau manifold also in geometric
terms and provides a link between the algebraic structure encoded
in $N=2$ minimal models and geometry of the manifold.
These mutually complementary approaches of CFT and LG are very efficient
also in investigation of $D$-branes on Calabi-Yau manifolds
~\cite{ReS}-~\cite{LiZ}.

 The free field realization of
the unitary $N=2$ super-Virasoro algebra representations has been
developed by Feigin and Semikhatov in ~\cite{FeS}. From the one
hand, it gives an efficient way for correlation functions calculation.
From the other hand it is closely related ~\cite{FST},
~\cite{FeS} with Wakimoto free field description of $SU(2)$ WZNW model
as well as with LG approach to $N=2$
minimal models. The effect of boundaries has already been
studed in LG approach ~\cite{HIV}, ~\cite{GJ} and in the context
of coset model ~\cite{MMS}. Thus, it is important to extend the
free field realization to the case of $N=2$ minimal models
with boundaries.

 In this note we extend the construction of ~\cite{FeS} to the
case of $N=2$ minimal models with  boundaries. In section 1,
we review the free field realization and butterfly resolution
~\cite{FeS} of Feigin and Semikhatov of the unitary modules in
$N=2$ minimal
models and obtain free field representation for the characters.
In section 2, we construct Ishibashi states of $A,B$-types
in Fock modules.
In section 3, we obtaine in an explicit form $A,B$-type Ishibashi state
for each irreducible $N=2$ minimal model module using superposition of
Ishibashi states of Fock modules from butterfly resolution.
The coefficients of the superposition are fixed by imposing
BRST invariance condition which is similar to that of the bulk
theory. At the end of the section free field construction
of boundary states is represented using Cardy's prescription.
In the last section we briefly discuss free field representation for
boundary correlations functions as well as some generalizations of the
construction.

\vskip 10pt
\centerline{\bf 1. Free-field realization of $N=2$ minimal models}
\centerline{\bf irreducible representations.}
\vskip 10pt

 In this section we briefly discuss free-field construction of
Feigin and Semikhatov ~\cite{FeS} of the irreducible modules
in $N=2$ superconformal minimal models.

\leftline{\bf 1.1. Free-field representations of $N=2$ super-Virasoro
algebra.}

  We introduce (in the left-moving sector) the free bosonic fields
$X(z), X^{*}(z)$ and free fermionic fields $\psi(z), \psi^{*}(z)$,
so that its singular OPE's are given by
\ber
X^{*}(z_{1})X(z_{2})=\ln(z_{12})+reg.,\nmb
\psi^{*}(z_{1})\psi(z_{2})=z_{12}^{-1}+reg,
\label{1.ope}
\enr
where $z_{12}=z_{1}-z_{2}$. Then for an arbitrary number
$\mu$ the currents of $N=2$ super-Virasoro
algebra are given by
\ber
G^{+}(z)=\psi^{*}(z)\d X(z) -{1\ov \mu} \d \psi^{*}(z), \
G^{-}(z)=\psi(z) \d X^{*}(z)-\d \psi(z), \nmb
J(z)=\psi^{*}(z)\psi(z)+{1\ov \mu}\d X^{*}(z)-\d X(z), \nmb
T(z)=\d X(z)\d X^{*}(z)+
{1\ov 2}(\d \psi^{*}(z)\psi(z)-\psi^{*}(z)\d \psi(z))-\nmb
{1\ov 2}(\d^{2} X(z)+{1\ov \mu}\d^{2} X^{*}(z)),
\label{1.min}
\enr
and the central charge is
\be
c=3(1-2{1\ov \mu}).
\label{1.cent}
\en

 As usual, the fermions in NS sector are expanded into
half-integer modes:
\ber
\psi(z)=\sum_{r\in 1/2+Z}\psi[r]z^{-{1\ov 2}-r},\
\psi^{*}(z)=\sum_{r\in 1/2+Z}\psi^{*}[r]z^{-{1\ov 2}-r},\nmb
G^{\pm}(z)=\sum_{r\in 1/2+Z}G^{\pm}[r]z^{-{3\ov 2}-r},
\label{1.NS}
\enr
and they are expanded into integer modes in R sector:
\ber
\psi(z)=\sum_{r\in Z}\psi[r]z^{-{1\ov 2}-r},\
\psi^{*}(z)=\sum_{r\in Z}\psi^{*}[r]z^{-{1\ov 2}-r},\nmb
G^{\pm}(z)=\sum_{r\in Z}G^{\pm}[r]z^{-{3\ov 2}-r}.
\label{1.R}
\enr
The bosons $X(z),X^{*}(z),J(z),T(z)$ are expanded in both sectors into integer
modes:
\ber
\d X(z)=\sum_{n\in Z}X[n]z^{-1-n},\
\d X^{*}(z)=\sum_{n\in Z}X^{*}[n]z^{-1-n},\nmb
J(z)=\sum_{n\in Z}J[n]z^{-1-n},\
T(z)=\sum_{n\in Z}L[n]z^{-2-n}.
\label{1.Bex}
\enr

 To describe the modules of $N=2$ Virasoro superalgebra in NS sector
we define the vacuum state $|p,p^{*}>$ such that
\ber
\psi[r]|p,p^{*}>=\psi^{*}[r]|p,p^{*}>=0, r\geq {1\ov 2},\nmb
X[n]|p,p^{*}>=X^{*}[n]|p,p^{*}>=0, n\geq 1, \nmb
X[0]|p,p^{*}>=p|p,p^{*}>, \
X^{*}[0]|p,p^{*}>=p^{*}|p,p^{*}>.
\label{1.vac}
\enr
It is a primary state with respect
to the $N=2$ Virasoro algebra
\ber
G^{\pm}[r]|p,p^{*}>=0, r>0, \nmb
J[n]|p,p^{*}>=L[n]|p,p^{*}>=0, n>0, \nmb
J[0]|p,p^{*}>={j\ov \mu}|p,p^{*}>=0, \nmb
L[0]|p,p^{*}>={h(h+2)-j^{2}\ov 4\mu}|p,p^{*}>=0,
\label{1.hwv}
\enr
where $j=p^{*}-\mu p$, $h=p^{*}+\mu p$. The vacuum state
$|p,p^{*}>$ corresponds to the vertex operator
$V_{(p,p^{*})}(z)\equiv\exp(pX^{*}(z)+p^{*}X(z))$ placed at $z=0$.

We denote by $F_{p,p^{*}}$
the Fock module generated from the vector $|p,p^{*}>$ by the fermionic
operators $\psi^{*}[r]$, $\psi[r]$, $r<{1\ov2}$, and bosonic operators
$X^{*}[n]$, $X[n]$, $n<0$.

 It is easy to
calculate the character $\om_{p,p^{*}}(q,u)$ of the Fock module
$F_{p,p^{*}}$. By the definition
\be
\om_{p,p^{*}}(q,u)=Tr_{F_{p,p^{*}}}(q^{L[0]-{c\ov 24}}u^{J[0]}).
\label{1.fchd}
\en
Thus we obtain
\be
\om_{p,p^{*}}(q,u)=
q^{{h(h+2)-j^{2}\ov 4\mu}-{c\ov24}}u^{{j\ov \mu}}
{\Theta(q,u)\ov \eta(q)^{3}},
\label{1.fchc}
\en
where we have used the Jacobi theta-function
\be
\Theta(q,u)=
q^{1\ov 8}\sum_{m\in Z}q^{{1\ov2}m^{2}}u^{-m}
\label{1.Jthet}
\en
and the Dedekind eta-function
\be
\eta(q)=q^{1\ov 24}\prod_{m=1}(1-q^{m}).
\label{1.eta}
\en

 The $N=2$ Virasoro algebra has the following set of automorphisms
which is known as spectral flow ~\cite{SS}
\ber
G^{\pm}[r]\rightarrow G_{t}^{\pm}[r]\equiv G^{\pm}[r\pm t], \nmb
L[n]\rightarrow L_{t}[n]\equiv L[n]+t J[n]+t^{2}{c\ov 6}\dlt_{n,0},\
J[n]\rightarrow J_{t}[n]\equiv J[n]+t {c\ov 3}\dlt_{n,0},
\label{1.flow}
\enr
where $t\in Z$. Note that spectral flow is intrinsic property
of $N=2$ super-Virasoro algebra and hence, it does not depend on a
particular realization. Allowing in (\ref{1.flow}) $t$ to be
half-integer, we obtain the isomorphism between the NS and R
sectors.

 The spectral flow action on the free
fields can be easily described if we bosonise fermions $\psi^{*}, \psi$
\be
\psi(z)=\exp(-y(z)), \ \psi^{*}(z)=\exp(+y(z)).
\label{1.fbos}
\en
and introduce spectral flow vertex operator ~\cite{LVW}
\be
U_{t}(z)=\exp(-t(y+{1\ov \mu}X^{*}-X)(z)).
\label{1.vflow}
\en
The following OPE's
\ber
\psi(z_{1})U_{t}(z_{2})=
z_{12}^{t}:\psi(z_{1})U_{t}(z_{2}):, \nmb
\psi^{*}(z_{1})U_{t}(z_{2})=
z_{12}^{-t}:\psi^{*}(z_{1})U_{t}(z_{2}):, \nmb
\d X^{*}(z_{1})U_{t}(z_{2})=z_{12}^{-1}t U_{t}(z_{2})+r.,\nmb
\d X(z_{1})U_{t}(z_{2})=-z_{12}^{-1}{t \ov \mu}U_{t}(z_{2})+r.
\label{1.2}
\enr
give the action of spectral flow on the modes of the free fields
\ber
\psi[r]\rightarrow \psi[r-t], \
\psi^{*}[r]\rightarrow \psi^{*}[r+t], \nmb
X^{*}[n]\rightarrow X^{*}[n]+t\dlt_{n,0}, \
X[n]\rightarrow X[n]-{t\ov \mu}\dlt_{n,0}.
\label{1.3}
\enr

 The action of the spectral flow on the vertex operator
$V_{(p,p^{*})}(z)$ is given by the normal ordered product of the
vertex $U_{t}(z)$ and $V_{p,p^{*}}(z)$.

\leftline{\bf 1.2. Irreducible $N=2$ super-Virasoro representations
and butterfly resolution.}

 The $N=2$ minimal models
are characterized by the condition that
$\mu$ is integer and $\mu\geq 2$. In NS sector the irreducible
highest-weight modules, constituting the (left-moving) space of states of
the minimal model, are unitary and labeled by two integers $h,j$,
where $h=0,...,\mu-2$ and $j=-h,-h+2,...,h$.
The highest-weight vector $|w_{h,j}>$ of the module satisfies the conditions
(which are similar to (\ref{1.hwv}))
\ber
G^{\pm}[r]|w_{h,j}>=0, r>0, \nmb
J[n]|w_{h,j}>=L[n]|w_{h,j}>=0, n>0, \nmb
J[0]|w_{h,j}>={j\ov \mu}|w_{h,j}>, \nmb
L[0]|w_{h,j}>={h(h+2)-j^{2}\ov 4\mu}|w_{h,j}>.
\label{1.hw}
\enr
If in addition to the conditions (\ref{1.hw}) the relation
\be
G^{+}[-1/2]|w_{h,j}>=0
\label{1.chw}
\en
is satisfied we call the vector $|w_{h,j}>$ and the module $M_{h,j}$
chiral highest-weight vector (chiral primary state) and chiral module,
correspondingly. In this case we have $h=j$.
Analogously, anti-chiral highest-weight vector (anti-chiral primary
state) and anti-chiral
module can be defined if instead of (\ref{1.chw})
\be
G^{-}[-1/2]|w_{h,j}>=0
\label{1.achw}
\en
is satisfied. In this case $h=-j$.
In ~\cite{FeS} the highest-weight vectors satisfying (\ref{1.hw})
are called massive highest-weight vectors and the vectors
satisfying in addition to (\ref{1.chw})
are called topological highest-weight vectors. In this paper we
prefer to use the terms introduced in ~\cite{LVW}.

 As we have seen in the preceding subsection, the highest
weight vectors (\ref{1.hw}) can be realized by the Fock vacuum
vectors $|p,p^{*}>$. But the corresponding Fock modules
are reducible with respect to $N=2$ super-Virasoro algebra.
To construct irreducible representations one needs to
introduce the integer lattice of the momentums:
\be
P=\{(p,p^{*})|p,p^{*}\in Z\}
\label{1.L}
\en
and the space
\be
F_{P}=\oplus_{(p,p^{*})\in P}F_{p,p^{*}}.
\label{1.fock}
\en
Following ~\cite{FeS} we introduce two fermionic screening currents
$S^{\pm}(z)$ and the charges $Q^{\pm}$ of the currents
\ber
S^{+}(z)=\psi^{*}\exp(X^{*})(z), \
S^{-}(z)=\psi\exp(\mu X)(z), \nmb
Q^{\pm}=\oint dz S^{\pm}(z)
\label{1.chrg}
\enr
These charges commute with the generators of $N=2$ super-Virasoro algebra
(\ref{1.min}) and act in the space $F_{P}$. Moreover they are
nilpotent and mutually anticommute
\be
(Q^{+})^{2}=(Q^{-})^{2}=\{Q^{+},Q^{-}\}=0.
\label{1.BRST}
\en
Due to these properties one can combine the charges $Q^{\pm}$ into
BRST operator acting in $F_{P}$ and build a BRST complex
consisting of Fock modules $F_{p,p^{*}}\in F_{P}$ such that its cohomology
is given by NS sector  $N=2$ minimal model irreducible module $M_{h,j}$.
This complex has been constructed in ~\cite{FeS}.

 Let us consider first free field construction
for the chiral module $M_{h,h}$. In this case the complex
(which is known due to Feigin and Semikhatov as butterfly resolution)
can be represented by the following diagram
\ber
\begin{array}{ccccccccccc}
&&\vdots &\vdots &&&&&&\\
&&\uparrow &\uparrow &&&&&&\\
\ldots &\leftarrow &F_{1,h+\mu} &\leftarrow
F_{0,h+\mu}&&&&&&\\
&&\uparrow &\uparrow &&&&&&\\
\ldots &\leftarrow &F_{1,h} &\leftarrow F_{0,h}&&&&&&\\
&&&&\nwarrow&&&&&\\
&&&&&F_{-1,h-\mu}&\leftarrow &F_{-2,h-\mu}&\leftarrow&\ldots\\
&&&&&\uparrow &&\uparrow&\\
&&&&&F_{-1,h-2\mu}&\leftarrow &F_{-2,h-2\mu}&\leftarrow&\ldots\\
&&&&&\uparrow &&\uparrow &&\\
&&&&&\vdots &&\vdots &&
\end{array} \nmb
\label{1.but}
\enr
The horizontal arrows in this diagram are given by the action of
$Q^{+}$ and vertical arrows are given by the action of $Q^{-}$.
The diagonal arrow at the middle of butterfly resolution
is given by the action of $Q^{+}Q^{-}$ (which equals $-Q^{-}Q^{+}$
due to (\ref{1.BRST})). Ghost number operator $g$ of this complex
is defined for an arbitrary vector $|v_{n,m}>\in
F_{n,m\mu+h}$ by
\ber
g|v_{n,m}>=(n+m)|v_{n,m}>, \ if \ n,m\geq 0, \nmb
g|v_{n,m}>=(n+m+1)|v_{n,m}>, \ if \ n,m< 0.
\label{1.grad}
\enr
For an arbitrary vector of the complex $|v_{N}>$ with the
ghost number $N$ the differential $d_{N}$ is defined
by
\ber
d_{N}|v_{N}>=(Q^{+}+Q^{-})|v_{N}>, \ if \ N\neq -1, \nmb
d_{N}|v_{N}>=Q^{+}Q^{-}|v_{N}>, \ if \ N=-1.
\label{1.diff}
\enr
and rises the ghost number by 1.

\leftline{\bf Theorem 1.1.}~\cite{FeS}. Complex (\ref{1.but}) is exact
except at the $F_{0,h}$ module, where the cohomology is given by
the chiral module $M_{h,h}$.

 The butterfly resolution allows to write the character
$\chi_{h}(q,u)\equiv Tr_{M_{h,h}}(q^{L[0]-{c\ov 24}}u^{J[0]})$ of the
module $M_{h,h}$ as an alternated sum:
\ber
\chi_{h}(q,u)=\chi^{(l)}_{h}(q,u)-\chi^{(r)}_{h}(q,u), \nmb
\chi^{(l)}_{h}(q,u)=\sum_{n,m\geq 0}(-1)^{n+m}\om_{n,h+m\mu}(q,u),\nmb
\chi^{(r)}_{h}(q,u)=\sum_{n,m> 0}(-1)^{n+m}\om_{-n,h-m\mu}(q,u),
\label{1.char}
\enr
where $\chi^{(l)}_{h}(q,u)$ and $\chi^{(r)}_{h}(q,u)$ are the characters
of the left and right wings of the resolution.

 To obtain the resolutions for other (anti-chiral and non-chiral) modules
we note first that all irreducible modules can be obtained
from the chiral modules $M_{h,h}$, $h=0,...,\mu-2$ by the spectral flow
action ~\cite{FeSST}. It turns out that one can get
all the resolutions by the spectral flow action also.
Indeed, the charges
$Q^{\pm}$ commute with spectral flow operator $U_{t}$ as it is
easy to see from the corresponding OPE's, hence, the resolutions
in NS sector are generated from (\ref{1.but}) by the operators
$U_{t}$, $t=-h,-h+1,...,-1$. The resolutions in R sector are
generated from the resolutions in NS sector by the spectral flow
operator $U_{-{1\ov2}}$.

 To illustrate how it works we consider the Fock module $F_{0,h}$ from
the middle of the resolution. The vector $|0,h>$, $(h>0)$ represents $N=2$
super-Virasoro algebra chiral highest-weight vector $|w_{h,h}>$
of the module $M_{h,h}$. The vector $\psi[-1/2]|0,h>=
{1\ov h}G^{-}[-1/2]|0,h>$ represents a cohomology class, hence
it is in irreducible module $M_{h,h}$. Acting on this vector by
the operator $U_{-1}$ we obtain the vector $|{1\ov \mu},h-1>$,
which is $N=2$ super-Virasoro algebra highest-weight vector $|w_{h,h-2}>$
of the module $M_{h,h-2}$. Analogously, the vector
$\psi[-3/2]\psi[-1/2]|0,h>={1\ov h(h-1)}G^{-}[-3/2]G^{-}[-1/2]|0,h>$
represents another vector from irreducible module $M_{h,h}$.
Acting on this vector by $U_{-2}$ we obtain the vector
$|{2\ov \mu},h-2>$, which is $N=2$ super-Virasoro algebra
highest-weight vector $|w_{h,h-4}>$ of the module $M_{h,h-4}$. Going
by this way further, we arrive at the end the vector
$|{h\ov \mu},0>=U_{-h}{1\ov h!}G^{-}[1/2-h]...G^{-}[-1/2]|0,h>$,
which is anti-chiral highest-weight vector $|w_{h,-h}>$ of the
anti-chiral module $M_{h,-h}$. The
modules from R sector can be generated analogously.

\vskip 10pt
\centerline{\bf 2. Ishibashi states in the Fock modules of $N=2$
super- Virasoro algebra.}

 In this section we begin to develop free field representation
of $N=2$ minimal models Ishibashi states.
Thus, it
will be implied in what follows that the right-moving sector of
the model is realized by the free-fields
$\bar{X}(\bar{z}), \bar{X}^{*}(\bar{z}), \bar{\psi}(\bar{z}),
\bar{\psi}^{*}(\bar{z})$, and the right-moving
$N=2$ super-Virasoro algebra is given by the formulas similar to
(\ref{1.min}).

 There are two types of boundary states preserving $N=2$
super-Virasoro algebra ~\cite{OOY}, usually called $B$-type
\ber
(L[n]-\bar{L}[-n])|B>>=(J[n]+\bar{J}[-n])|B>>=0, \nmb
(G^{+}[r]+\imath \et \bar{G}^{+}[-r])|B>>=
(G^{-}[r]+\imath \et \bar{G}^{-}[-r])|B>>=0
\label{2.BD}
\enr
and $A$-type states
\ber
(L[n]-\bar{L}[-n])|A>>=(J[n]-\bar{J}[-n])|A>>=0, \nmb
(G^{+}[r]+\im \et \bar{G}^{-}[-r])|A>>=
(G^{-}[r]+\im \et \bar{G}^{+}[-r])|A>>=0
\label{2.AD}
\enr
where $\et=\pm 1$.
The Ishibashi states (as well as the
boundary states) can be considered
as the linear functionals on the space of states of the $N=2$
minimal model. From the other hand,
we have seen in the Sec.1, that the space of states in the left-
moving sector is represented by the cohomology groups
of butterfly resolutions. It is clear that similar construction
can be applied for the right-moving space of states.
Therefore, free field
construction of Ishibashi states has to be consistent with the
resolutions. This problem of consistence will be postponed to the next
section. In this section we consider the most simple solutions of
(\ref{2.BD}), (\ref{2.AD}) in the tensor product
of the left-moving Fock module $F_{p,p^{*}}$ and right-moving Fock
module $\bar{F}_{\bar{p},\bar{p}^{*}}$. We shall call these states as
linear Ishibashi~\cite{Ish}
states and denote by $|p,p^{*},\bar{p},\bar{p}^{*},\et,B(A)>>$.

 Let us consider $B$-type linear Ishibashi states in NS sector.
They can be easily obtained from the following ansatz for fermions
\ber
(\psi^{*}[r]-
\im a\bar{\psi}^{*}[-r])|p,p^{*},\bar{p},\bar{p}^{*},\et,B>>=0, \nmb
(\psi[r]-
\im b\bar{\psi}[-r])|p,p^{*},\bar{p},\bar{p}^{*},\et,B>>=0
\label{2.Bantz}
\enr
where $a,b$ are the arbitrary constants. Substituting these relations
into (\ref{2.BD}) and using (\ref{1.min}) we find
\ber
a=b=\et, \nmb
\bar{p}=-p-{1\ov \mu}, \ \bar{p}^{*}=-p^{*}-1, \nmb
(X[n]+\bar{X}[-n]+{1\ov \mu}\dlt_{n,0})
|p,p^{*},\bar{p},\bar{p}^{*},\et,B>>=0, \nmb
(X^{*}[n]+\bar{X}^{*}[-n]+\dlt_{n,0})
|p,p^{*},\bar{p},\bar{p}^{*},\et,B>>=0.
\label{2.BX}
\enr
Thus, the linear $B$-type Ishibashi state in NS sector is given by the
standard expression ~\cite{FPSLR}, ~\cite{CLNY},~\cite{PC}
\ber
|p,p^{*},\et,B>>=
\prod_{n=1}\exp(-{1\ov
n}(X^{*}[-n]\bar{X}[-n]+X[-n]\bar{X}^{*}[-n])) \nmb
\prod_{r=1/2}\exp(\im\et(\psi^{*}[-r]\bar{\psi}[-r]+
\psi[-r]\bar{\psi}^{*}[-r]))|p,p^{*},-p-{1\ov \mu},-p^{*}-1>.
\label{2.Bish}
\enr
The closed string cylinder amplitude between such states in NS sector
is given by
\ber
<<p_{2},p^{*}_{2},\et,B|q^{L[0]-c/24}u^{J[0]}|p_{1},p^{*}_{1},\et,B>>=\nmb
\dlt(p_{1}-p_{2})
\dlt(p^{*}_{1}-p^{*}_{2})\om_{p_{1},p^{*}_{1}}(q,u).
\label{2.ampl}
\enr

Note that the state $<<p,p^{*},\et,B|$ is defined in such
a way to satisfy conjugate boundary conditions and to take into account
the charge asymmetry ~\cite{FeFu}-~\cite{Fel}
of the free-field realization of the minimal model.

 Introducing the new set of bosonic oscillators
\ber
v[n]={\im\ov\sqrt{2\mu}}(X^{*}[n]-\mu X[n]), \nmb
u[n]={1\ov\sqrt{2\mu}}(X^{*}[n]+\mu X[n]),
\label{2.tor}
\enr
one can rewrite the $B$-type conditions (\ref{2.BX}) as
\ber
(v[n]+\bar{v}[-n])|p,p^{*},\bar{p},\bar{p}^{*},\et,B>>=0, \nmb
(u[n]+\bar{u}[-n]+
\sqrt{{2\ov\mu}}\dlt_{n,0})|p,p^{*},\bar{p},\bar{p}^{*},\et,B>>=0.
\label{2.NN}
\enr

 One can consider the
coordinates $(\exp(u),v)$ as the polar coordinates on a complex
plane and think of the conditions (\ref{2.NN}) as Neumann
along both of coordinates $(u,v)$.
From this point of view the bosonic part
$\im\sqrt{{2\ov \mu}}(v[0]+\bar{v}[0])$
of the sum $J[0]+\bar{J}[0]$ generates an action of a circle on the
complex plane such that $v$ is an angular coordinate.
Then, $B$-type states correspond to 2-dimensional $D$-branes
filling the complex plane. One can equally well to think of the
conditions (\ref{2.NN}) as Dirichlet ones along $(u,v)$ such that
the circle action on the complex plane is generated by
$\im\sqrt{{2\ov \mu}}(v[0]-\bar{v}[0])$ and then, $B$-type boundary
states correspond to $D0$-branes. There are also two additional
interpretations of the conditions when we consider one of the
relations (\ref{2.NN}) as Neumann and another one as Dirichlet
condition. Thus, one has four possibilities. As we shall see at the
end of this section, mirror symmetry left only two of them

 The linear $A$-type Ishibashi states can be found analogously.
We start from the ansatz for fermions
\ber
(\psi^{*}[r]-
\im a\bar{\psi}[-r])|p,p^{*},\bar{p},\bar{p}^{*},\et,A>>=0, \nmb
(\psi[r]-
\im b\bar{\psi}^{*}[-r])|p,p^{*},\bar{p},\bar{p}^{*},\et,A>>=0
\label{2.Aantz}
\enr
where $a,b$ are the arbitrary constants. Then we find
\ber
a=\et \mu, \ b={\et\ov \mu}, \nmb
\bar{p}=-{1+p^{*}\ov \mu}, \ \bar{p}^{*}=-\mu p-1, \nmb
(\mu X[n]+\bar{X}^{*}[-n]+\dlt_{n,0})
|p,p^{*},\bar{p},\bar{p}^{*},\et,A>>=0, \nmb
(X^{*}[n]+\mu\bar{X}[-n]+\dlt_{n,0})
|p,p^{*},\bar{p},\bar{p}^{*},\et,A>>=0.
\label{2.AX}
\enr
Hence the linear $A$-type Ishibashi state (in NS sector) is given by
\ber
|p,p^{*},\et,A>>=
\prod_{n=1}\exp(-{1\ov
n}(\mu X[-n]\bar{X}[-n]+{1\ov \mu}X^{*}[-n]\bar{X}^{*}[-n])) \nmb
\prod_{r=1/2}\exp(\im\et({1\ov \mu}\psi^{*}[-r]\bar{\psi}^{*}[-r]+
\mu\psi[-r]\bar{\psi}[-r]))|p,p^{*},-{1+p^{*}\ov \mu},-\mu p-1>
\label{2.Aish}
\enr
and the corresponding closed string cylinder amplitude
coincide with the (\ref{2.ampl}).

 One can find that (\ref{2.AX}) can be rewritten in $(u,v)$
coordinates as
\ber
(v[n]-\bar{v}[-n])|p,p^{*},\bar{p},\bar{p}^{*},\et,A>>=0, \nmb
(u[n]+\bar{u}[-n]+
\sqrt{{2\ov\mu}}\dlt_{n,0})|p,p^{*},\bar{p},\bar{p}^{*},\et,A>>=0.
\label{B.DN}
\enr
One can consider these relations as Neumann condition along
the coordinate $u$ and Dirichlet condition along $v$.
Then, $A$-type states correspond to 1-dimensional
$D$-branes along the rays in complex plane which is in agreement with
the results ~\cite{HIV}, ~\cite{GJ} obtained in LG approach.
But we are free to choose, similar to the case of $B$-type states,
three other possible interpretations.
Because of $A$-type Ishibashi state (\ref{2.Aish}) can be obtained
from $B$-type Ishibashi state (\ref{2.Bish}) by the Mirror
involution $\sgm$ in the right-moving sector:
\ber
\sgm \bar{v}[n]=-\bar{v}[n], \
\sgm \bar{u}[n]=\bar{u}[n], \nmb
\sgm\bar{\psi}[r]={1\ov\mu}\bar{\psi}^{*}[r], \
\sgm\bar{\psi}^{*}[r]=\mu\bar{\psi}[r]
\label{2.mirr}
\enr
we are left with two possibilities for $A$-type states:
$D1$-branes along the rays or along the circles of the complex
plane, and we have two possibilities for $B$-type states:
$D2$-branes or $D0$-branes. But, Poincare duality relates to each other
the states for each type, so that we are left with $D1$-brane for
$A$-type and $D0$-brane for $B$-type.

\vskip 10pt
\centerline{\bf 3. Boundary states in $N=2$ minimal models.}
\vskip 10pt
\leftline{\bf 3.1. Ishibashi states in the irreducible modules of $N=2$
super-Virasoro algebra.}

 In this section we represent free field construction of Ishibashi
states of irreducible modules $M_{h,j}$. The construction uses
the linear Ishibashi states (\ref{2.Bish}), (\ref{2.Aish}) as the
building blocks in such a way to be consistent with butterfly
resolutions of irreducible modules.

 The relations
(\ref{1.char}), (\ref{2.ampl}) indicate that Ishibashi state of
irreducible module has to be given as a superposition of
linear Ishibashi states of the Fock modules.
Indeed, let us consider the following
superposition of $B$-type free-field Ishibashi states
\be
|M_{h,h},\et,B>>=\sum_{n,m\geq0}c_{n,m}|n,m\mu+h,\et,B>>+
\sum_{n,m>0}c_{-n,-m}|-n,-m\mu+h,\et,B>>,
\label{2.MIB}
\en
where the summation is performed over the momentums
from the butterfly resolution (\ref{1.but}). It is easy to see that
this state satisfies the relations (\ref{2.BD}). The arbitrary
coefficients $c_{n,m}$ and $c_{-n,-m}$ can be fixed partly
from the condition that closed string cylinder amplitude between
the Ishibashi states gives the characters of irreducible modules.
In the free field realization it is equivalent to the relation
\be
<<M_{h',h'},\et,B|(-1)^{g}q^{L[0]-c/24}u^{J[0]}|M_{h,h},\et,B>>=
\dlt_{h',h}\chi_{h}(q,u).
\label{2.MBampl}
\en
Indeed, using (\ref{2.ampl}) one can find
\ber
<<M_{h',h'},\et,B|(-1)^{g}q^{L[0]-c/24}u^{J[0]}|M_{h,h},\et,B>>=\nmb
\dlt_{h',h}(\sum_{n,m\geq0}(-1)^{n+m}|c_{n,m}|^{2}\om_{n,h+m\mu}(q,u)-
\sum_{n,m>0}(-1)^{n+m}|c_{-n,-m}|^{2}\om_{-n,h-m\mu}(q,u)).
\label{2.cnorm}
\enr
Comparing with (\ref{1.char}) we find
\be
|c_{n,m}|^{2}=|c_{-n,-m}|^{2}=1.
\label{2.cmod}
\en
Thus, the state (\ref{2.MIB}) is a good candidate for free-field
realization (in NS sector) of $B$-type Ishibashi state of the chiral
module $M_{h,h}$. It would be a genuine Ishibashi state if it did
not radiate nonphysical closed string states which are present in the
free field representation of the model. In other words, the overlap of
this state with an
arbitrary closed string state which does not belong to
the Hilbert space of the $N=2$ minimal model should vanish. As we will see
this condition can be formulated as a $BRST$ invariance condition
of the state (\ref{2.MIB}) and it will fix the coefficients
$c_{n,m}$, $c_{-n,-m}$ up to the common factor.

 To formulate $BRST$ invariance condition one has to consider 
what kind of closed string states
in the free field realization can interract with the state
(\ref{2.MIB}). They come from the product of left-moving and right-moving
Fock modules
$F_{n,h+m\mu}\otimes\bar{F}_{-n-{1\ov\mu},-1-h-m\mu}$, where
$n,m\geq0$ or $n,m<0$. The left-moving modules of the superposition
(\ref{2.MIB}) constitute the
butterfly resolution (\ref{1.but}) whose cohomology is given by
the irreducible chiral module $M_{h,h}$. What about the Fock modules from
the right-moving sector? They don't form the resolution
like (\ref{1.but}) due to the relations for the momentums
from (\ref{2.BX}) and (\ref{2.AX}). Instead the
right-moving Fock modules constitute the dual butterfly resolution:
\ber
\begin{array}{ccccccccccc}
&&\vdots &\vdots &&&&&&\\
&&\downarrow &\downarrow &&&&&&\\
\ldots &\rightarrow &\bar{F}_{-1-{1\ov\mu},-1-h-\mu} &\rightarrow
\bar{F}_{-{1\ov\mu},-1-h-\mu}&&&&&&\\
&&\downarrow &\downarrow &&&&&&\\
\ldots &\rightarrow &\bar{F}_{-1-{1\ov\mu},-1-h} &\rightarrow
\bar{F}_{-{1\ov\mu},-1-h}&&&&&&\\
&&&&\searrow&&&&&\\
&&&&&\bar{F}_{1-{1\ov\mu},-1-h+\mu}&\rightarrow &\bar{F}_{2-{1\ov\mu},-1-h+\mu}
&\rightarrow&\ldots\\
&&&&&\downarrow &&\downarrow&\\
&&&&&\bar{F}_{1-{1\ov\mu},-1-h+2\mu}&\rightarrow &\bar{F}_{2-{1\ov\mu},-1-h+2\mu}
&\rightarrow&\ldots\\
&&&&&\downarrow &&\downarrow &&\\
&&&&&\vdots &&\vdots &&
\end{array} \nmb
\label{2.dualbut}
\enr
The arrows on this diagram are given by the same operators as on
the diagram (\ref{1.but}). We define the ghost number operator
$\bar{g}$ of the complex similar to the ghost number operator of the
complex (\ref{1.but}). For an arbitrary vector $|\bar{v}_{n,m}>\in
\bar{F}_{n-{1\ov\mu},m\mu-1-h}$ by
\ber
\bar{g}|\bar{v}_{n,m}>=(n+m)|v_{n,m}>, \ if \ n,m\leq 0, \nmb
\bar{g}|v_{n,m}>=(n+m-1)|v_{n,m}>, \ if \ n,m> 0.
\label{2.grad}
\enr
The differentials $\bar{d}_{\bar{N}}$ are defined similar to
(\ref{1.diff}). This resolution can also be used for description
of irreducible modules due to

\leftline{\bf Theorem 3.1.} Complex (\ref{2.dualbut}) is exact
except at the $\bar{F}_{-{1\ov\mu},-1-h}$ module, where the cohomology is
given by the anti-chiral module $M_{h,-h}$.

 The proof of this theorem is similar to that one is given in
~\cite{FeS} for the complex (\ref{1.but}).

Thus, the states which can interact with (\ref{2.MIB}) come from
the product of resolution (\ref{1.but}) and (\ref{2.dualbut}).
The tensor product of complexes (\ref{1.but}) and
(\ref{2.dualbut}) constitutes the complex
\be
\ldots\rightarrow C_{h,h}^{-2}\rightarrow C_{h,h}^{-1}\rightarrow
C_{h,h}^{0}\rightarrow C_{h,h}^{+1}\rightarrow\ldots,
\label{2.complex}
\en
which is graded by the sum of the ghost numbers
$g+\bar{g}$ and for an arbitrary ghost number $i$
the space $C_{h,h}^{i}$ is given by the sum of products of the
Fock modules from the resolution (\ref{1.but}) and (\ref{2.dualbut})
such that $g+\bar{g}=i$. The differential $D$ of the complex
(\ref{2.complex}) is defined by the differentials $d_{N}$ and
$\bar{d_{\bar{N}}}$ of the complexes (\ref{1.but}) and (\ref{2.dualbut})
\be
D_{i}|v_{N}\otimes\bar{v}_{\bar{N}}>=|d_{N}v_{N}\otimes\bar{v}_{\bar{N}}>+
(-1)^{N}|v_{N}\otimes\bar{d}_{\bar{N}}\bar{v}_{\bar{N}}>,
\label{2.Diff}
\en
where $|v_{N}>$ is an arbitrary vector from the complex (\ref{1.but})
with ghost number $N$, while $|\bar{v}_{\bar{N}}>$ is an
arbitrary vector from the complex (\ref{2.dualbut}) with the ghost
number $\bar{N}$ and $N+\bar{N}=i$.
It follows from the Theorems (1.1) and (3.1) that the cohomology
${\bf H}^{*}$
of the complex (\ref{2.complex}) is nonzero only at grading 0
and it is given by the product of irreducible modules
$M_{h,h}\otimes\bar{M}_{h,-h}$.

 The Ishibashi state we are looking for
can be considerd as a linear functional on the Hilbert space of
$N=2$ superconformal minimal model, then it has to be an element
of the homology group ${\bf H}_{*}$. Therefore, the $BRST$ invariance
condition for the state can be formulated as follows.

 Let us define the action of the differential $D$ on the state
$|M_{h,h},\et,B>>$ by the relation
\be
<<D^{*}M_{h,h},\et,B|v_{N}\otimes \bar{v}_{\bar{N}}>\equiv
<<M_{h,h},\et,B|D_{N+\bar{N}}|v_{N}\otimes \bar{v}_{\bar{N}}>,
\label{2.D*}
\en
where $v_{N}\otimes \bar{v}_{\bar{N}}$ is an arbitrary element
from $C_{h,h}^{N+\bar{N}}$. Then, $BRST$ invariance condition
means that
\be
D^{*}|M_{h,h},\et,B>>=0.
\label{2.Dcycl}
\en

\leftline{\bf Theorem 3.2.} Superposition (\ref{2.MIB}) satisfies
$BRST$ invariance condition (\ref{2.Dcycl}) if the coefficients
$c_{n,m}$, $c_{-n,-m}$ of the superposition obey the equations
\ber
c_{n,m}=c_{0,0}, \ if \ n+m=2k, \nmb
c_{n,m}=-\im\et c_{0,0}, \ if \ n+m=2k+1.
\label{2.cycle}
\enr
Thus the coefficients depends on the ghost number $g$.

 Proof. In view of (\ref{2.MIB}, \ref{2.Bish}) and because of
differential $D$ rises the ghost number by 1, only $BRST$ images of the
states $|v_{N}\otimes\bar{v}_{-1-N}>\in C_{h,h}^{-1}$ have nonzero overlap
with Ishibashi state $|M_{h,h},\et,B>>$. Thus, one needs to show that
\ber
<<M_{h,h},\et,B|D_{-1}|v_{N}\otimes\bar{v}_{-1-N}>= \nmb
<<M_{h,h},\et,B|d_{N}|v_{N}\otimes\bar{v}_{-1-N}>
+(-1)^{N}<<M_{h,h},\et,B| \bar{d}_{-1-N}|v_{N}\otimes\bar{v}_{-1-N}>=0.
\label{2.eq}
\enr
It will be implied during the proof that the state
$D_{-1}|v_{N}\otimes\bar{v}_{-1-N}>$ corresponds to the field
$(D_{-1}(v_{N}\otimes\bar{v}_{-1-N}))(z,\bar{z})$ which is placed
at the center $z=\bar{z}=0$ of the unit disk.

 1)Let $N\geq0$ and hence $v_{N}$ and $\bar{v}_{-1-N}$ belong to the left
wings of resolutions (\ref{1.but}) and (\ref{2.dualbut}). Using
relations (\ref{2.Bantz}) and (\ref{2.BX}) as well as the
definitions (\ref{1.diff}), (\ref{1.chrg}) one can rewrite the
first term from (\ref{2.eq}) as
\ber
<<M_{h,h},\et,B|d_{N}|v_{N}\otimes\bar{v}_{-1-N}>=\nmb
\im \et<<M_{h,h},\et,B|(\exp(X^{*}_{0}-\bar{X}^{*}_{0})\bar{Q}^{+}+
\exp(\mu(X_{0}-\bar{X}_{0}))\bar{Q}^{-})|v_{N}\otimes\bar{v}_{-1-N}>,
\label{2.term1}
\enr
where $X^{*}_{0}$, $\bar{X}^{*}_{0}$, $X_{0}$, $\bar{X}_{0}$ are
the constant modes canonicaly conjugated to the momentums
$X[0]$, $\bar{X}[0]$, $X^{*}[0]$, $\bar{X}^{*}[0]$. The operators
$\exp(X^{*}_{0}-\bar{X}^{*}_{0})$, $\exp(X_{0}-\bar{X}_{0})$
only shift the momentums in the components of the superposition
$<<M_{h,h},\et,B|$. From the other hand we can decompose the vectors
$|v_{N}>$ and $|\bar{v}_{-1-N}>$ according to their components
in the Fock modules of the resolution
\ber
|v_{N}>=|y_{N,h}>+|y_{N-1,h+\mu}>+...+|y_{0,h+N\mu}>,\nmb
|\bar{v}_{-1-N}>=|\bar{y}_{-1-N-{1\ov\mu},-1-h}>+
|\bar{y}_{-N-{1\ov\mu},-1-h+\mu}>+...
+|\bar{y}_{-{1\ov\mu},-1-h+(N+1)\mu}>
\label{2.dec}
\enr
Substitution of this decomposition into the each term of the equation
(\ref{2.eq}) gives (\ref{2.cycle}).

 2)Let $N=-1$ and hence $v_{-1}$ and $\bar{v}_{0}$ belong to the
vertexes of the right wings of resolutions. Using
relations (\ref{2.Bantz}) and (\ref{2.BX}) as well as the
definitions (\ref{1.diff}), (\ref{1.chrg}) one can rewrite the
first term from (\ref{2.eq}) as
\ber
<<M_{h,h},\et,B|d_{-1}|v_{-1}\otimes\bar{v}_{0}>=\nmb
<<M_{h,h},\et,B|\exp(X^{*}_{0}-\bar{X}^{*}_{0})\exp(\mu(X_{0}-\bar{X}_{0}))
\bar{Q}^{+}\bar{Q}^{-}|v_{-1}\otimes\bar{v}_{0}>.
\label{2.term2}
\enr
Substitution of this decomposition into the each term of the equation
(\ref{2.eq}) gives the relation from
(\ref{2.cycle}) when $n=m=-1$.

 3)Let $N<-1$ and hence $v_{N}$ and $\bar{v}_{-1-N}$ belong to the
right wings of resolutions. This case can be treated similar to the
case $N\geq0$ and hence we obtain (\ref{2.cycle}). It proves the
theorem.

 Note that $BRST$-closed state (\ref{2.MIB}, \ref{2.cycle}) is not
$BRST$-exact due to (\ref{2.MBampl}).
Hence, it represents a homology class from $\bf{H_{*}}$ and it is defined
modulo $BRST$-exact states satisfying (\ref{2.BD}). Note also that
normalization phase $c_{0,0}$ can not be
fixed by the $BRST$ invariance condition. We fix the normalization of
each Isibashi state by $c_{0,0}=1$.

 Due to the arguments from the Sec.1 we can
obtain free field construction of the remainder $N=2$ minimal model
$B$-type Ishibashi states in NS sector applying to the Ishibashi state
$|M_{h,h},\et,B>>$ the spectral flow operators $U_{t}\bar{U}_{-t}$, where
$t=-h,-h+1,...-1$. Then, the action by the spectral flow operator
$U_{-{1\ov2}}\bar{U}_{{1\ov2}}$ on the Ishibashi
states from the NS sector gives free field construction of
$B$-type Ishibashi states in R sector. It is also clear that
free field construction of $A$-type Ishibashi states in $N=2$
minimal models can be obtained from $B$-type Ishibashi states
by the Mirror involution (\ref{2.mirr}).

\leftline{\bf 3.2. Boundary states.}

 Free field representation of the boundary states in NS or R sector can be
constructed by applying Cardy's prescription ~\cite{C} to free field
realized Ishibashi states.

 Let us denote by $S_{(h,j),(h',j')}$ the S-matrix of modualr
transformation of the full characters of $N=2$ minimal model in NS sector:
\ber
\chi_{h,j}(q,0)=\sum_{h',j'}S_{(h,j),(h',j')}\chi_{h',j'}(\tld{q},0),\nmb
S_{(h,j),(h',j')}=
{1\ov\sqrt{2}\mu}\sin({\pi(h+1)(h'+1)\ov\mu})\exp(\im\pi{jj'\ov\mu}).
\label{2.smat}
\enr
where $q=\exp(\im 2\pi\tau)$, $\tld{q}=\exp(-\im {2\pi\ov \tau})$.
Then Cardy's formula in NS sector gives the following boundary states
\be
|D_{h,j},\et,A>>=
\sum_{h',j'}D_{(h,j),(h',j')}|M_{h',j'},\et,A>>,
\label{2.NScardy}
\en
while in R sector it gives
\be
|D_{h,j},\pm,A>>=
\pm\im\sum_{(h',j')}D_{(h,j),(h',j')}U_{1\ov2}\bar{U}_{1\ov2}
|M_{h',j'},\pm,A>>,
\label{2.Rcardy}
\en
where
\be
D_{(h,j),(h',j')}={S_{(h,j),(h',j')}\ov \sqrt{S_{(0,0),(h',j')}}}
\label{2.Dcoef}
\en
and $h'=0,...,\mu-2$, $j'=-h',-h'+2,...,h'$.

 Free field $B$-type boundary states can be obtained from $A$-type
boundary states by the orbifold projection ~\cite{MMS}. In
NS sector we obtain
\be
|D_{h,j},\et,B>>=
\sum_{h'}D_{(h,j),(h',0)}|M_{h',0},\et,B>>,
\label{2.NSB}
\en
while in R sector we have the following
\be
|D_{h,j},\pm,B>>=
\pm\im\sum_{h'}D_{(h,j),(h',0)}U_{1\ov2}\bar{U}_{-1\ov2}
|M_{h',0},\pm,B>>,
\label{2.RB}
\en
where $h'=0,...,\mu-2$.

The Ishibashi states in the expressions (\ref{2.NScardy}), (\ref{2.Rcardy}),
(\ref{2.NSB}), (\ref{2.RB}) correspond to the full
characters of the $N=2$ Virasoro superalgebra. Free field realization
of the
standard $A(B)$-type Ishibashi states $|h,j,s,A(B)>>$, where $s=0,2$ in NS sector
and $s=-1,1$ in R sector is given by the relations
\ber
|M_{h,j},+,A(B)>>=|h,j,0,A(B)>>+|h,j,2,A(B)>>, \nmb
|M_{h,j},-,A(B)>>=|h,j,0,A(B)>>-|h,j,2,A(B)>>,
\label{2.NSstand}
\enr
in NS sector and it is given by
\ber
U_{1\ov2}\bar{U}_{1\ov2}|M_{h,j},+,A(B)>>=|h,j,-1,A(B)>>+|h,j,1,A(B)>>, \nmb
U_{1\ov2}\bar{U}_{1\ov2}|M_{h,j},-,A(B)>>=|h,j,-1,A(B)>>-|h,j,1,A(B)>>,
\label{2.Rstand}
\enr
in R sector.

 Due to the brief discussion of boundary conditions in Sec.2
we have the following geometric interpretation of boundary states
in $N=2$ minimal model. $A$-type boundary states can be considered
as $D1$-branes along the rays or along the circles in the complex plane
and they are Poincare dual to each other. This is in agreement
with the results ~\cite{HIV}, ~\cite{GJ} obtained in LG approach.
$B$-type boundary states can be considered as $D0$ or $D2$-branes
which are also Poincare dual to each other.

\vskip 10pt
\centerline{\bf4. Discussion}

 In this note we represented free field construction
of Ishibashi and boundary states
in $N=2$ superconformal minimal models using free field realization
of $N=2$ super-Virasoro algebra unitary modules.
Each Ishibashi state of the model is given by
infinite superposition of linear Ishibashi states of Fock modules,
forming butterfly resolution of irreducible representation
of $N=2$ super-Virasoro algebra. It is shown that
coefficients of the superposition are fixed by the $BRST$ invariance condition
and the Ishibashi state constructed this way is $BRST$ closed but
not $BRST$ exact and represents thereby a homology class.
We group these free field realized Ishibashi
states into the boundary states of $N=2$ minimal model using the
solution found by Cardy. Due to $BRST$ invariance the boundary states
do not radiate non-physical closed string states (which are present
originally in the free field space of states). We found that
$B$-type boundary
states corresponds to $D0$ or $D2$-branes in complex plane
target space. $A$-type D-branes is natural to identify with rays
or circles in complex plane. This identification is in agreement with
the results ~\cite{HIV}, ~\cite{GJ} obtained in LG approach
but more detail investigation of
D-brane geometry in the free field approach needs to be done.
It would be interesting to find geometric interpretation of
the superposition of linear Ishibashi states involved into the free
field construction of the $D$-branes. In the R sector it is a
brane anti-brane system due to (\ref{2.cycle}) and it would be
interesting to find the interpretation of this superposition in the
context of tachyon condensation ~\cite{Sen}.

 We close with a brief discussion of some
directions to develop. At first we would like to point out
that our free field construction can be easily generalized
to the case of orientifolds. The second problem is a free field
construction for boundary correlation functions. This problem is
rather technical in fact because all needed ingredients are known.
Indeed, free field representation of the boundary states is obtained
in the present note. The free field description of the irreducible
representations, vertex operators and the screening charge generating
quantum group structure of the conformal blocks (note that the fermionic
screenings $Q^{\pm}$ of the butterfly resolution have trivial
braiding relations) are known from ~\cite{FeS}. The
screening charge is
nothing else the standard Wakimoto screening charge (the one
involved in the costruction of Felder-type resolution
~\cite{BFeld}) of $SU(2)$ WZNW model. In terms of the free fields it
is given by the integral of the (bosonic) screening current:
\be
Q_{W}=\oint dz S_{W}(z), \
S_{W}=
(\d X^{*}+\psi^{*}\psi)\exp(-{1\ov\mu}X^{*}-X).
\label{4.1}
\en
It is easy to check that $Q_{W}$ commutes with $N=2$
super-Virasoro algebra and fermionic charges $Q^{\pm}$.
We hope to develop free field representation of the boundary
correlation functions in future publication.

 The third interesting direction is the generalization of
our boundary state construction to the case of
nondiagonal modular invariant partition functions.
It is obviously important to generalize the free field
construction of boundary states to the case of
Gepner models.

 Free field approach is also applies to $N=2$ superconformal
models with $W$-algebra of symmetries. Some of them is believed
to coincide with Kazama-Suzuki models ~\cite{Ito}. $N=2$ minimal models
is the simplest example of this situation when Kazama-Suzuki model is
$SU(2)\times U(1)/U(1)$ coset and
it would be interesting to extend free field construction of
$D$-branes to more general class of Kazama-Suzuki models.

\vskip 10pt
\centerline{\bf Acknowledgements}
I am grateful to B.L.Feigin, A.M.Semikhatov and I.Yu.Tipunin for
very useful discussions.

 This work was supported in part by grants RBRF-01-02-16686,
RBRF-00-15-96579, INTAS-00-00055
.

\end{document}